\documentclass[letterpaper, english, twocolumn, aps]{revtex4-1}
\usepackage{textcomp}
\usepackage{amsmath}
\usepackage{amssymb}
\usepackage{graphicx}
\usepackage[francais,english]{babel}
\usepackage[utf8]{inputenc}
\usepackage{hyperref}
\usepackage{wasysym}
\usepackage{graphics}
\usepackage{color}


\usepackage{babel}
\begin{document}

\preprint{PRE}

\title{Convergence in nonlinear laser wakefield accelerators modeling in a Lorentz-boosted frame}

\author{P. Lee}
\email{patrick.lee@u-psud.fr}
\affiliation{LPGP, CNRS, Univ. Paris-Sud, Université Paris-Saclay, 91405, Orsay, France}

\author{J.-L.Vay}
\affiliation{Lawrence Berkeley National Laboratory, Berkeley, CA 94720, USA}

\date{\today}
\begin{abstract}
Laser wakefield acceleration modeling using the Lorentz-boosted frame technique in the particle-in-cell code has demonstrated orders of magnitude speedups. A convergence study was previously conducted in cases with external injection in the linear regime and without injection in the nonlinear regime, and the obtained results have shown a convergence within the percentage level. In this article, a convergence study is carried out to model electron self-injection in the 2-1/2D configuration. It is observed that the Lorentz-boosted frame technique is capable of modeling complex particle dynamics with a significant speedup. This result is crucial to curtail the computational time of the modeling of future chains of $10\,\mathrm{GeV}$ laser wakefield accelerator stages with high accuracy.

\end{abstract}


\maketitle
\section{Introduction}
Electron acceleration via Laser Wakefield Acceleration (LWFA) relies on the interaction between an intense laser pulse and an underdense plasma to generate a plasma wave that can support a high accelerating gradient, typically of the order of $\mathrm{GV/m}$ \cite{tajima_laser_1979, esarey_physics_2009, malka_laser_2012}. This mechanism allows for the production of femtoseconds-length electrons beams with GeV energy that are applicable to various domains, such as in ultrafast electron diffraction, radiography, or, in the future, using chains of tens of plasma acceleration stages to accelerate electrons and positrons to 1 TeV energy or more to answer fundamental questions regarding e.g. the origins of the universe or of dark energy.

The Particle-In-Cell (PIC) algorithm has been the method of choice for numerical modeling of LWFA experiments. The most commonly used electromagnetic formulation uses second-order finite-difference discretization of Maxwell's equations in both space and time, known also as Finite-Difference Time-Dependent (FDTD) or Yee solver. This formulation allows fast resolution and good scaling in parallel, but suffers from various anomalous numerical effects resulting from discretization, such as numerical dispersion. To improve the efficiency and accuracy of the Yee solver, Non-Standard Finite-Difference (NSFD) solvers were introduced. Among them is the ``Cole-Karkkainnen'' (CK) \cite{cole_high-accuracy_1997,cole_high-accuracy_2002} solver, which enlarges the stencil in the direction transverse to the finite differencing, thereby allowing a larger time step than with the standard Yee solver. In addition, the CK solver does not have numerical dispersion along the principal axes at the Courant-Friedrich-Lewy (CFL) limit \cite{courant_partial_1967} for a given time step and parameters, provided that the cell size is the same along each dimension, i.e. cubic cells in 3D, or along the shortest cell for an appropriate choice of parameters \cite{cowan_generalized_2013}. To eliminate altogether the numerical dispersion, Haber et al. introduced a pseudo-spectral analytical time-domain (PSATD) algorithm \cite{haber_advances_1973}, which has no CFL limit, offers substantial flexibility in plasma and particle beam simulations, and is more stable with regard to Numerical Cerenkov instability \cite{godfrey_improved_2015}.

Computer simulations of LWFA experiments using the PIC algorithm require to resolve the evolution of a laser driver and an accelerated particle beam into a plasma structure that is of orders of magnitude longer and wider than the accelerated beam. The laser wavelength is usually on the scale of $1\,\mathrm{\mu m}$ while the length of the plasma structure can be on the scale of $1$ to $10^{3}\,\mathrm{mm}$. This disparity in cell size and propagation distance results in very computationally intensive simulations. Furthermore, laser power and energy are increasing at the time of writing, allowing beam energies beyond $10\,\mathrm{GeV}$ in the next decade to be attained using longer plasma structures or chains of plasma stages, requiring more computational resources. To scale up with this, several approaches may be considered such as simulations with reduced model \cite{cowan_characteristics_2011,benedetti_efficient_2012,mora_kinetic_1997}, advances in high performance computing \cite{vincenti_efficient_2017}, or simulations with the Lorentz-boosted frame technique \cite{vay_noninvariance_2007}, a method that can curtail computational time by several orders of magnitude. The focus of this article is on the latter.

The Lorentz-boosted frame technique \cite{vay_noninvariance_2007} relies on the use of a frame of reference moving at relativistic velocity with regard to the laboratory frame, leading to space-time Lorentz contraction and dilation of the experimental components. In LWFA, the  scale  gap  between  the  laser  pulse  and  the  plasma  structure  can  be  reduced  by  choosing  an  optimal  frame  of  reference  that  travels  close  to  the  speed  of  light  in  the  direction  of  the  laser  pulse. In such a frame, the laser pulse wavelength increases, and the plasma length decreases, while at the same time, the time scale of the response of the laser pulse to the plasma decreases, and the time scale of the response of the plasma to the laser increases. Matching the spatial and temporal scales leads to gains, as the crossing time between the laser pulse and the plasma column is reduced. The choice of the optimal frame is guided by the spatial and time resolutions required to capture the relevant physics in a given frame, and thus depends on the specific setup under consideration.

Several studies have been carried out on the accuracy of the Lorentz boosted frame technique. LWFA simulations with external injection \cite{vay_effects_2011} of electron beam in the linear wakefield was previously studied and the results on the evolution of the laser and electron beam properties have a 99\% agreement between simulations using various reference frames. In \cite{yu_enabling_2016}, the author has studied the convergence of the evolution of the laser between the laboratory (lab) frame in quasi-3D geometry and in the boosted frame. The results that were reported demonstrated good agreement in the blowout regime and without self-injection, however some discrepancies are observed in the case with self-injection of electrons, which involves strong nonlinear particle dynamics. In \cite{martins_numerical_2010, yu_enabling_2016}, the authors underlined that more accurate results can be obtained with a high number of macro-particles in the injected bunch to allow for significant statistics. 

In this article, we report on a convergence study of simulations using the Lorentz-boosted frame technique with CK and PSATD solvers. The outcome of this study shows that the Lorentz-boosted frame technique retains the accuracy in the modeling of self-injection in the nonlinear regime at high resolution, while obtaining significant speedups, and reports, for the first time, convergence at the percent level in the nonlinear regime with self-injection. Convergence is demonstrated on the electron bunch charge and energy, and also on energy spread and emittance, which are more sensitive to the numerical resolution \cite{lehe_laser-plasma_2014}.
The rest of the article is presented as follows. In Section~\ref{sec:setup}, the simulation setups in the boosted frame are discussed. Section~\ref{sec:results} shows results obtained in two case studies at different plasma densities: $\mathrm{10^{19}\,cm^{-3}}$ in Section~\ref{sec:high_dens}, and $\mathrm{10^{18}\,cm^{-3}}$ in Section~\ref{sec:low_dens}. The study is completed with a runtime analysis in Section~\ref{sec:runtime}. 

\section{Simulation setups in the boosted frame}
\label{sec:setup}
This section presents the modeling of the dynamics of the self-injected electrons in the blowout regime in 2-1/2D using the Lorentz-boosted frame technique implemented in Warp \cite{vay_novel_2012}.

The main physical and numerical parameters of the simulations are given in Table~\ref{table:list-parameters}. They were chosen to be close (though not identical) to a case reported in \cite{cormier-michel_scaled_2009,vay_modeling_2010}, with the main difference being the value of $a_0 = 5$ at $n_e=10^{19}\,\mathrm{cm^{-3}}$, and $a_0 = 8$ at $n_e=10^{18}\,\mathrm{cm^{-3}}$. The high value of $a_0$ was chosen to trigger wavebreaking, a necessary condition for electron self-injection in the wakefield in order to study its dynamics. The simulations were performed for stages accelerating to a few tens of $\mathrm{MeV}$ using a plasma density of $10^{19}\,\mathrm{cm}^{-3}$, and close to a $\mathrm{GeV}$ using a plasma density of $10^{18}\,\mathrm{cm}^{-3}$. The latter is one of the configurations that is being considered as the first stage in the EuPRAXIA project \cite{eupraxia_compact_nodate}. These simulations are run using both the CK (using Cowan's parameter settings \cite{cowan_generalized_2013}) and the PSATD solvers, and with a $4$-pass bilinear filter plus compensation \cite{vay_numerical_2011}.

\begin{table}[htb]
\centering
\caption{List of parameters for a LWFA electron injector simulation}
\label{table:list-parameters}
\resizebox{\columnwidth}{!}{%
\begin{tabular}{lll}
\hline
\hline
\multicolumn{1}{l}{Plasma density on axis} & \multicolumn{1}{c}{$n_0$} &  \multicolumn{1}{c}{[$10^{18}, 10^{19}]\,\mathrm{cm^{-3}}$} \\
\multicolumn{1}{l}{Plasma longitudinal profile} & \multicolumn{1}{c}{} & \multicolumn{1}{c}{Entrance ramp + plateau} \\
\multicolumn{1}{l}{Plasma transverse profile} & \multicolumn{1}{c}{} & \multicolumn{1}{c}{Uniform} \\
Plasma length & \multicolumn{1}{c}{$L_{plasma}$} & \multicolumn{1}{c}{$[1,0.05]\,\mathrm{cm}$} \\
Plasma entrance ramp profile &  & \multicolumn{1}{c}{linear} \\
Plasma entrance ramp length &  & \multicolumn{1}{c}{$[150$, $50]\,\mathrm{\mu m}$} \\
 &  &  \\
Laser profile &  & \multicolumn{1}{c}{\protect\footnotemark bi-Gaussian} \\
Laser polarization &  & \multicolumn{1}{c}{linear (in $y-$direction)} \\
Laser focal position &  \multicolumn{1}{c}{$z_f$} & \multicolumn{1}{c}{$0\,\mathrm{mm}$} \\
Peak normalized laser field strength & \multicolumn{1}{c}{$a_0(z_f)$} & \multicolumn{1}{c}{$[8,\, 5]$} \\
Laser wavelength & \multicolumn{1}{c}{$\lambda_0$} & \multicolumn{1}{c}{$0.8\,\mathrm{\mu m}$} \\
Normalized laser spot size & \multicolumn{1}{c}{$k_p\sigma$} & \multicolumn{1}{c}{$5.3$} \\
Normalized laser length & \multicolumn{1}{c}{$k_pL$} & \multicolumn{1}{c}{$2$} \\
 &  &  \\
Boundary conditions & & \begin{tabular}[c]{@{}c@{}} \multicolumn{1}{c}{Open boundaries in}\\
\multicolumn{1}{c}{$x-,\,z-$directions with PML}\end{tabular}\\
Stencil order (for PSATD solver) &  & \multicolumn{1}{c}{32} \\
Cell size in x &  & \multicolumn{1}{c}{$[0.33, 0.1\,\mathrm{\mu m}]$} \\
Cell size in z &  & \multicolumn{1}{c}{$\lambda_0/128-\lambda_0/16$} \\
Time-step &  &  \multicolumn{1}{c}{At the CFL limit}\\
Particle deposition order &  & \multicolumn{1}{c}{Cubic} \\
Number of plasma particles/cell &  & \multicolumn{1}{c}{$4\times 4\,$(in $x-,\,z-$directions)} \\ 
\hline
\hline
\end{tabular}%
}
\begin{flushleft}
$^{a}$Gaussian in temporal and spatial profiles
\end{flushleft}
\end{table}

The laser group velocities evaluated for the given parameters using the linear plasma fluid theory are $\gamma_g \approx 13.2$, and $41.8$ for $10^{19}\,\mathrm{cm^{-3}}$ and $10^{18}\,\mathrm{cm^{-3}}$ respectively. Note that for $a_0$ of $5$ and $8$, as used here, the group velocity of the wake is smaller than the one given by linear theory. Indeed, the presence of strongly nonlinear effects in this regime, such as self-compression or self-focusing of the laser pulse put a constraint on the choice of $\gamma_b$. In this regard, $\gamma_b$ cannot be given directly by the laser group velocity predicted by the linear plasma fluid theory, however using a heuristic approach and measurements from existing simulations, $\gamma_b$ was estimated to be $\approx 0.25\gamma_g$, with $\gamma_g$ predicted by the linear plasma fluid theory. The high density case with $10^{19}\,\mathrm{cm^{-3}}$ was first investigated. 
Warp simulations were performed for $\gamma_b$ between $1$ and $3$ and for longitudinal resolutions ranging from $N_z/\lambda_0 = 16$ to $N_z/\lambda_0 = 128$. Note that $\gamma_b=1$ is the lab frame. The same study approach was carried out at lower density, at $10^{18}\,\mathrm{cm^{-3}}$, for which $\gamma_b$ of $6,\,10$ were considered. 

The physical features observed in the boosted frame are somewhat different from the ones in the lab frame, in accordance with the properties of the Lorentz transformation \cite{vay_numerical_2011,vay_modeling_2010}, rendering direct comparison difficult. Thus to enable comparison between simulations with different $\gamma_b$, an inverse Lorentz transformation is performed to convert boosted frame data back to the laboratory frame. The reconstruction of the laboratory frame data from the boosted frame data is similar to those described in \cite{vay_application_2009, martins_exploring_2010}.

\section{Results}
\label{sec:results}

In this section, the results take the beam loaded longitudinal electric field, and the accelerated electron beam properties as figures of merit. Two case studies were considered: few tens of $\mathrm{MeV}$-stage at a plasma density of $\mathrm{10^{19}\,cm^{-3}}$ in Section~\ref{sec:high_dens}, and a nearly 1 $\mathrm{GeV}$-stage at $\mathrm{10^{18}\,cm^{-3}}$ in Section~\ref{sec:low_dens}.

\subsection{Plasma density at $10^{19}\,\mathrm{cm^{-3}}$}

Simulations with CK and PSATD solvers were conducted for the parameters shown in Table~\ref{table:list-parameters}. At $\mathrm{10^{19}\,cm^{-3}}$, the chosen relativistic boost factors are $\gamma_b\in[2,3]$, additional simulations with $\gamma_b=1$ were conducted for the CK solver. For each $\gamma_b$, a sweep of the longitudinal resolution, $N_z/\lambda_0$ from $16$ to $128$ was carried out. 
\label{sec:high_dens}

\subsubsection{Cole-Karkkainen solver}
\label{sec:high-CK}

We first evaluate results from simulations using the CK solver. 

\begin{figure}[htb]
\centering
\includegraphics{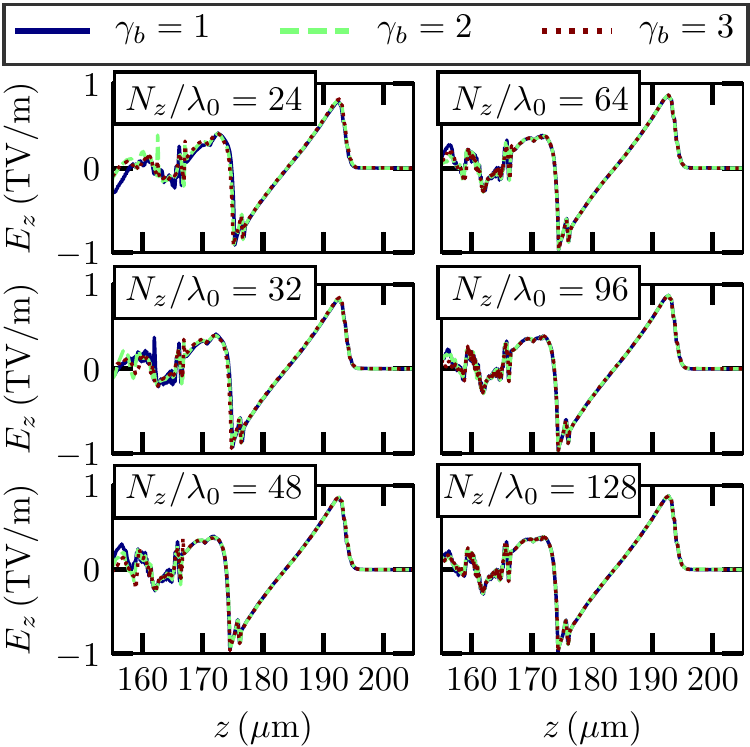}
\caption[A series of plots showing wakefield when the laser approaches $z=200\,\mathrm{\mu m}$ for the CK solver.]{
A series of plots showing wakefield at $z=200\,\mathrm{\mu m}$. Each panel corresponds to a specific longitudinal resolution given in the box on the upper left corner. Each panel shows the wakefield on axis from 2-1/2D simulations using the CK solver, carried out with $\gamma_b\in [1,2,3]$. Note that $\gamma_b=1$ represents the simulation in the lab frame.}
\label{fig:field-layout-FDTD}
\end{figure}

\begin{figure*}[htb]
\centering
\includegraphics{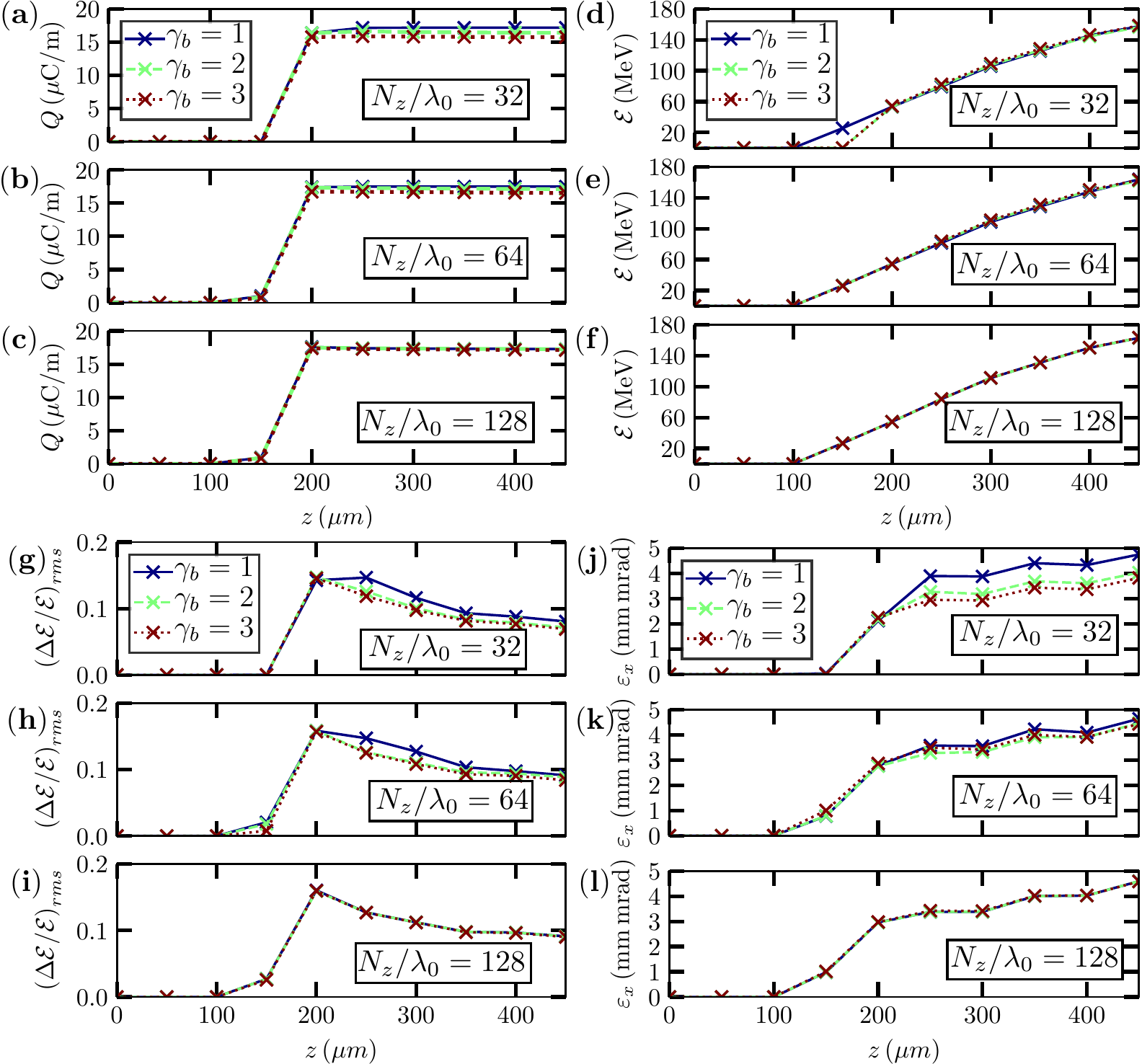}
\caption[Evolution of the injected electron bunch properties with respect to $z$, the distance of propagation in the plasma for the CK solver.]{Evolution of the injected electron bunch properties with respect to $z$, the distance of propagation in the plasma. Each plot illustrates simulations using different $\gamma_b\in[1,2,3]$ at a specific longitudinal resolution, $N_z/\lambda_0 \in[32,64,128]$ as indicated by the legend. Evolution of the electron bunch charge $Q$ is shown in (a-c); the average energy $\left<\mathcal{E}\right>$ in (d-f); the rms energy spread $(\Delta\mathcal{E}/\mathcal{E})_{rms}$ in (g-i); transverse emittance $\varepsilon_x$ in (j-l).}
\label{fig:beam-prop-evol-fdtd}
\end{figure*}

Fig.~\ref{fig:field-layout-FDTD} shows a layout of the wakefield $E_z$ on axis, captured when the laser approaches $z=200\,\mathrm{\mu m}$. Each panel corresponds to a specific longitudinal resolution. Wakefield simulations carried out with $\gamma_b\in [1,2,3]$ are compared in each panel. Results show some discrepancies in the wakefield at various resolutions, especially at the back of the first blow-out structure at $N_z/\lambda_0\leq48$. However, we observe a convergence with resolution of the wakefield for $N_z/\lambda_0>48$, and at $N_z/\lambda_0=128$, an excellent agreement is obtained for all $\gamma_b$ . The effect of beam loading is also visible for $N_z/\lambda_0>24$ at $z=175\,\mathrm{\mu m}$, confirming that the amplitude and the phase of beam loading are correctly recovered. This is further confirmed by the plot of the evolution of the injected bunch properties with respect to the propagation distance, $z$ in the lab frame, as described next (see Fig.~\ref{fig:beam-prop-evol-fdtd}). 

The evolution of the injected and accelerated electron bunch was also evaluated. Here we only consider electrons trapped in the first-period plasma wave (or first bucket). The selection of the evaluated electron bunch is detailed in Appendix. 
Fig.~\ref{fig:beam-prop-evol-fdtd} shows the evolution of the electron bunch properties as it propagates through the plasma for $\gamma_b\in[1,2,3]$. For each electron bunch property, the results are shown for varying longitudinal resolution $N_z/\lambda_0\in[32,64,128]$. From Fig.~\ref{fig:beam-prop-evol-fdtd}(a-c), it is observed that the injection happens from $z=100$ to $200\,\mu m$. For $z>200\,\mu m$, the electron bunch charge remains constant, implying that self-injection of electrons in the first plasma period has ended. The evolution of the average electron bunch energy is shown in Fig.~\ref{fig:beam-prop-evol-fdtd}(d-f). Once the electron bunch is injected, it is accelerated throughout the plasma to an average energy of $160\,\mathrm{MeV}$ at $z=450\,\mu m$. The average accelerating field $\left<{E_z}\right>$ is estimated at $5.3\,\mathrm{GeV/cm}$. The evolution of the rms energy spread $(\Delta \mathcal{E}/\mathcal{E})_{rms}$ of the electron bunch, shown in Fig.~\ref{fig:beam-prop-evol-fdtd}(g-i), suggests that it first reaches a maximum value at $z=200\,\mathrm{\mu m}$, then decreases, due to the increase of the average energy of the electron bunch, and plateaus at $\sim10\%$. Fig.~\ref{fig:beam-prop-evol-fdtd}(j-l) shows the evolution of the transverse emittance $\varepsilon_x$ of the injected electron bunch. A rapid emittance growth is observed during the injection, due to the plasma electrons circular motion in the wakefield cavity before being injected, thus gaining transverse momentum along the trajectory in the self-injection scheme. As a result, the non-zero transverse momentum contributes to the rapid emittance growth. Once the injection phase is over, the emittance growth slows down. This slow growth can be explained by the betatron movement of electrons. Since the betatron frequency depends on the energy of the individual electron, they do not all oscillate synchronously, resulting in the slow growth of the emittance. $\varepsilon_x$ reaches $\approx4.5\,\mathrm{mm\,mrad}$ at $z=450\mathrm{\mu m}$.

Fig.~\ref{fig:beam-prop-evol-fdtd} exhibits some discrepancies between results given by simulations with $N_z/\lambda_0\in [32,64]$ especially for second-order beam properties such as the energy spread and the emittance. In particular, the transverse emittance at $N_z/\lambda_0=32$ (Fig.~\ref{fig:beam-prop-evol-fdtd}(b)), we observe a difference of the order of $\sim10\%$ between $\gamma_b=1$ and $\gamma_b=3$. This indicates that the longitudinal resolution at $N_z/\lambda_0=32$ might not be sufficient to provide accurate modeling of the emittance. On the contrary, a nice agreement is observed for $N_z/\lambda_0 = 128$, suggesting that the higher the longitudinal resolution, the better the agreement between results from simulations with different $\gamma_b$. A convergence analysis is provided further in this section to enable quantitative comparison. 

For further and more thorough studies, an analysis centered on a specific frame (at $z=200\mathrm{\mu m}$) was conducted. Since results in Fig.~\ref{fig:beam-prop-evol-fdtd} have shown that all electron bunch properties are modeled correctly at all distances of propagation, $z$, for the highest longitudinal resolution $N_z/\lambda_0=128$, the choice of the frame is therefore unimportant.

\begin{figure}[htb]
\centering
\includegraphics{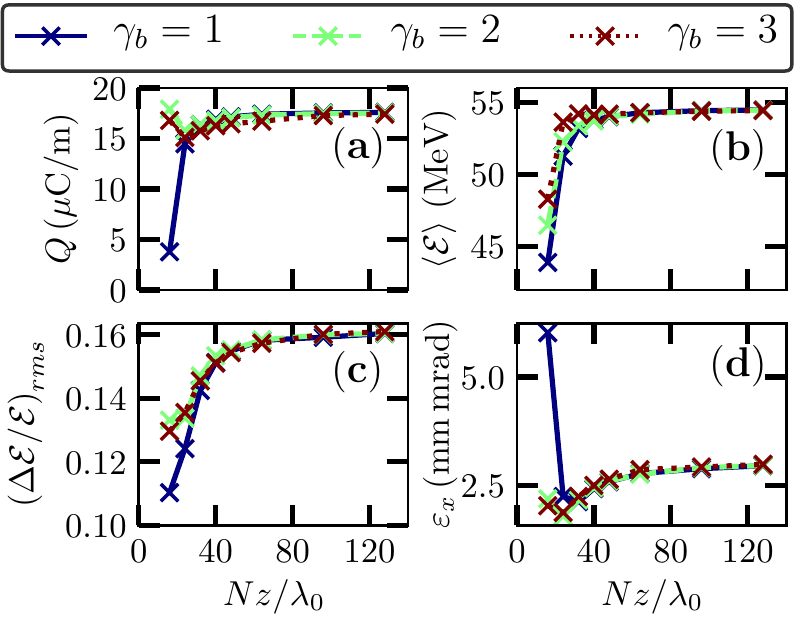}
\caption[Properties of injected and accelerated electron bunch evaluated at $z=200\,\mu m$ with respect to the longitudinal resolution for the CK solver.]{Properties of injected and accelerated electron bunch evaluated at $z=200\,\mu m$ with respect to the longitudinal resolution for $\gamma_b\in[1,2,3]$. Simulations were carried out using Warp in 2-1/2D using the CK solver. Note that $\gamma_b=1$ corresponds to the simulation in the lab frame. (a) Electron bunch charge, $Q$, (b) average energy, $\left<\mathcal{E}\right>$, (c) rms energy spread, $(\Delta \mathcal{E}/\mathcal{E})_{rms}$ and (d) tranverse emittance with respect to longitudinal resolution, $N_z/\lambda_0$ are illustrated.}
\label{fig:beam-prop-fdtd}
\end{figure}

Fig.~\ref{fig:beam-prop-fdtd} shows the injected and accelerated electron bunch (a) charge, (b) average energy, (c) rms energy spread, and (d) transverse emittance at frame $z=200\mathrm{\mu m}$ with respect to the longitudinal resolution, $N_z/\lambda_0$. Each plot shows results from simulations with $\gamma_b\in[1,2,3]$. 
There is a convergence of results obtained from simulations with different $\gamma_b$ for all electron bunch properties. 


\begin{figure}[htb]
\centering
\includegraphics[width=8.5cm]{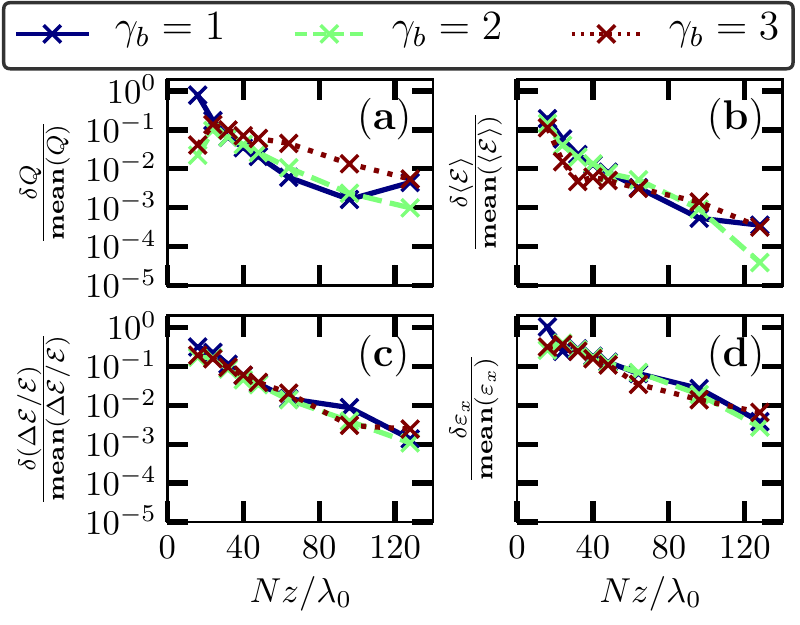}
\caption[Convergence analysis of the results obtained from simulations using the Lorentz-boosted frame technique for the CK solver.]{Convergence analysis of the results obtained from simulations using the Lorentz-boosted frame technique. The reference case is taken as the average of all considered relativistic factors. Each plot corresponds to different electron bunch properties: (a) the difference in electron bunch charge $\delta Q/\mathbf{mean}(Q)$; (b) in average energy,  $\delta \left<\mathcal{E}\right>/\mathbf{mean}(\left<\mathcal{E}\right>)$; (c) in rms energy spread, $\delta (\Delta\mathcal{E}/\mathcal{E})/\mathbf{mean}(\Delta\mathcal{E}/\mathcal{E})$; (d) in transverse emittance, $\delta\varepsilon_x/\mathbf{mean}(\varepsilon_x)$.}
\label{fig:beam-prop-fdtd-error}
\end{figure}

For a finer analysis, a quantification of the difference in convergence among all simulations of considered $\gamma_b$ has also been done. This convergence analysis takes the average of all considered relativistic factors at $N_z/\lambda_0=128$ as the reference case. This choice is made based on the fact that a convergence for all beam properties is attained at this resolution as shown in Fig.~\ref{fig:beam-prop-fdtd}. Fig.~\ref{fig:beam-prop-fdtd-error}(a-d) show the difference for each electron bunch properties represented in log scale in the y-axis with respect to $N_z/\lambda_0$. We observe that the difference in beam quantities decreases with respect to the resolution, confirming that high longitudinal resolution helps in attaining convergence. 
Notice that the rate of convergence is independent of $\gamma_b$, e.g. results from the lab frame do not converge faster than results in Lorentz-boosted frames. 
For $\gamma_b=1$ (lab frame) and $\gamma_b=2$, the difference is less than $1\%$ when $N_z/\lambda_0\geq64$ for all bunch properties except the beam emittance where a higher resolution ($N_z/\lambda_0=128$) is required to attain this difference margin. As for $\gamma_b=3$, the difference drops to less than $1\%$ for $N_z/\lambda_0\geq64$ for electron bunch average energy and rms energy spread, however a higher resolution ($N_z/\lambda_0=128$) is required to attain this difference margin for electron bunch charge and transverse emittance.  

These results demonstrated subpercent level convergence and confirm that increasing the resolution helps the convergence for all reference frames that were considered. In addition to the study of convergence with the longitudinal resolution, we have verified that increasing the transverse resolution also helps the convergence.

\subsubsection{PSATD solver}
\label{sec:high-PSATD}

Simulations with the laser-plasma parameters given in Table \ref{table:list-parameters} were carried out using the PSATD solver in 2-1/2D. The stencil of pseudo-spectral solvers can emulate a finite-difference stencil of arbitrary order \cite{vincenti_detailed_2016}. For this study, the stencil order was set to $32$. Since we are mostly interested in demonstrating convergence for simulations in a boosted frame, we have only performed simulations using the Lorentz-boosted frame technique with $\gamma_b\in[2,3]$ (simulations in the lab frame were not performed). The study approach is the same as for the CK solver in Section~\ref{sec:high-CK}. A sweep of longitudinal resolution was conducted for each relativistic factor of the boosted frame.

\begin{figure}[htb]
\centering
\includegraphics{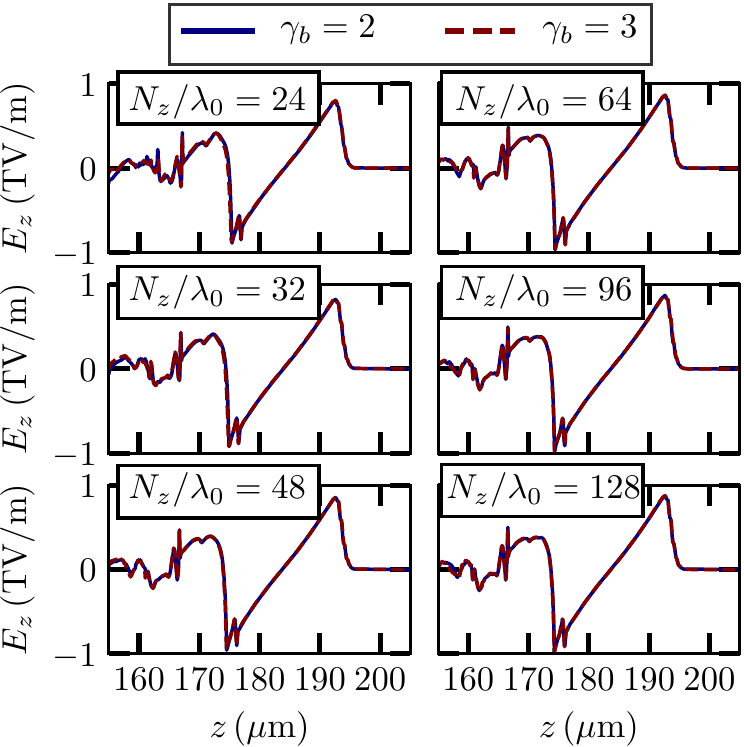}
\caption[A series of plots showing wakefield when the laser approaches $z=200\,\mathrm{\mu m}$ for the PSATD solver.]{Same as Fig.~\ref{fig:field-layout-FDTD} but with the PSATD solver.}
\label{fig:field-layout-spectral}
\end{figure}

\begin{figure*}[htb]
\centering
\includegraphics{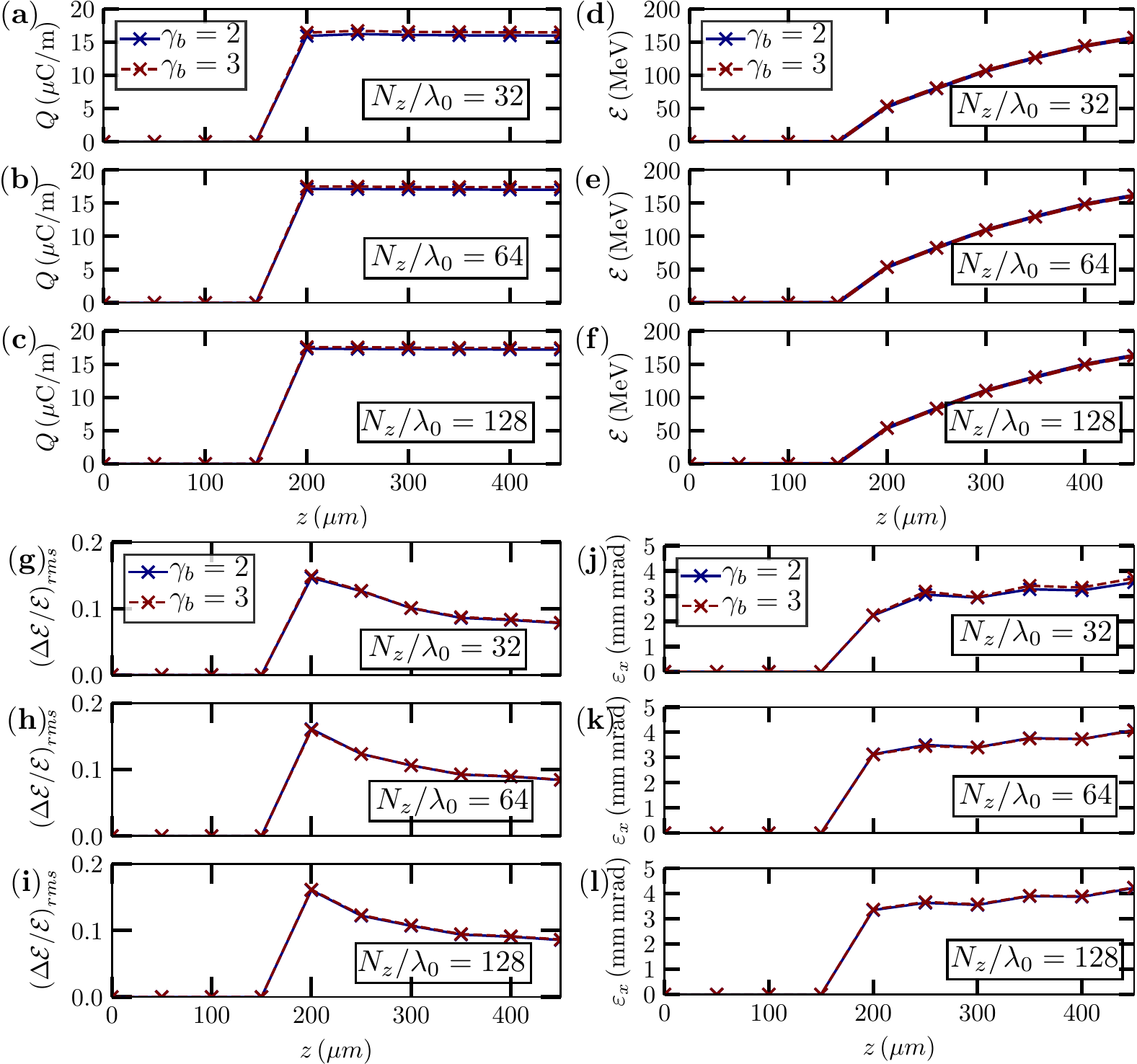}
\caption[Evolution of the injected electron bunch properties with respect to $z$, the distance of propagation in the plasma for the PSATD solver.]{Same as Fig.~\ref{fig:beam-prop-evol-fdtd} but with the PSATD solver.}
\label{fig:beam-spectral-evol}
\end{figure*}

The evaluation of the wakefield when the laser approaches $z=200\,\mathrm{\mu m}$ is reported in Fig.~\ref{fig:field-layout-spectral}. The wakefield from simulations with $\gamma_b\in[2,3]$ is illustrated in each plot for a specific longitudinal resolution. For $N_z/\lambda_0\geq32$ onwards, all wakefield structures for boosted frames with $\gamma_b\in[2,3]$ look identical. The beam loading effects at $z=175\mathrm{\mu m}$ are also captured in simulations in the Lorentz-boosted frame. The validity of the beam loading effects will be further confirmed by the injected electron bunch properties in the following paragraphs. 

Fig.~\ref{fig:beam-spectral-evol} shows the evolution of the injected electron bunch properties for several longitudinal resolutions, $N_z/\lambda_0\in[32,64,128]$. These bunch properties are electron bunch charge, average energy, rms energy spread and transverse emittance represented by Fig.~\ref{fig:beam-spectral-evol}(a-d) respectively. The injected electron bunch has a charge of $17.5\,\mathrm{\mu C}$, an average energy of $160\,\mathrm{MeV}$, a rms energy spread of $\sim10\%$ and a transverse emittance of $4.2\,\mathrm{mm\,mrad}$ at $z=450\,\mathrm{\mu m}$, which are comparable to the ones obtained using the CK solver, verifying convergence between the CK and the PSATD solvers. 

\begin{figure}[htb]
\centering
\includegraphics{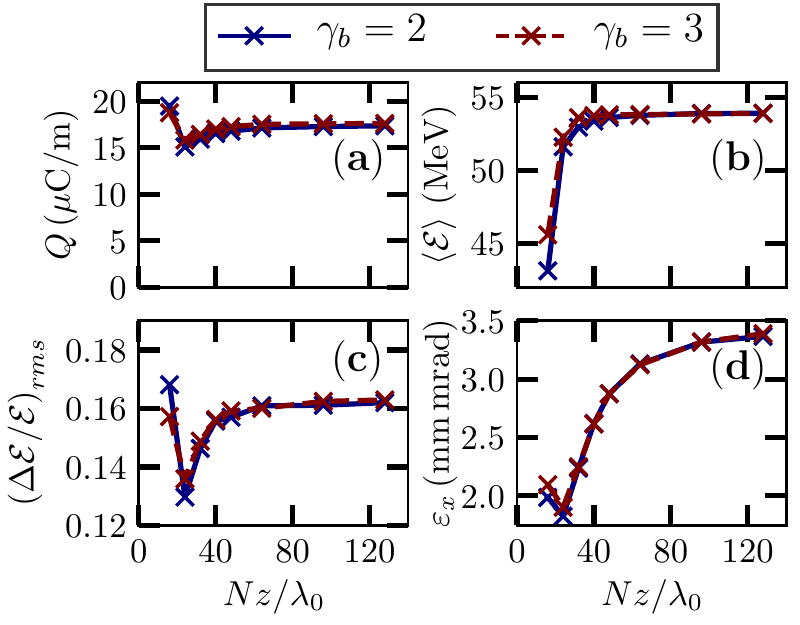}
\caption[Properties of injected and accelerated electron bunch for the PSATD solver using the Lorentz-boosted frame technique for the PSATD solver.]{Same as Fig.~\ref{fig:beam-prop-fdtd} but with the PSATD solver.}
\label{fig:beam-prop-spectral}
\end{figure}

The excellent agreement of the electron bunch properties with respect to the distance of propagation $z$ in Fig.~\ref{fig:beam-spectral-evol}, for $N_z/\lambda_0\in[32,64,128]$, allows us to further our analysis by looking in detailed results from a specific frame, $z=200\,\mathrm{\mu m}$. Fig.~\ref{fig:beam-prop-spectral} shows the electron bunch properties at $z=200\,\mathrm{\mu m}$ with respect to $N_z/\lambda_0$ for various $\gamma_b$. 
For all electron bunch properties, 
we observe a convergence of results for both $\gamma_b$ from $N_z/\lambda_0=48$ onwards.

\subsection{Plasma density at $10^{18}\,\mathrm{cm^{-3}}$}
\label{sec:low_dens}

The plasma density currently being explored for a laser-plasma injector is of the order of $10^{18}\,\mathrm{cm^{-3}}$ \cite{lee_dynamics_2016,lee_optimization_2017,brijesh_tuning_2012,burza_laser_2013,clayton_self-guided_2010,audet_investigation_2016,fourmaux_quasi-monoenergetic_2012,faure_injection_2010,kalmykov_electron_2009,pak_injection_2010}
At such density, with an intense laser pulse of $10^{18}\,\mathrm{W/cm^2}$, the generated wake has a large amplitude but is rather slow, creating favorable conditions for electron injection. In addition, the electron dephasing length of the order of $\mathrm{cm}$ scale allows electrons to be accelerated to the GeV range. Simulations using the laboratory frame at the aforementioned plasma density are rather impractical, therefore only simulations in boosted frames up to $\gamma_b \sim 0.25\gamma_g = 10$ are considered.

\subsubsection{Cole-Karkkainen solver}
\label{sec:low-CK}

\begin{figure}[htb]
\centering
\includegraphics{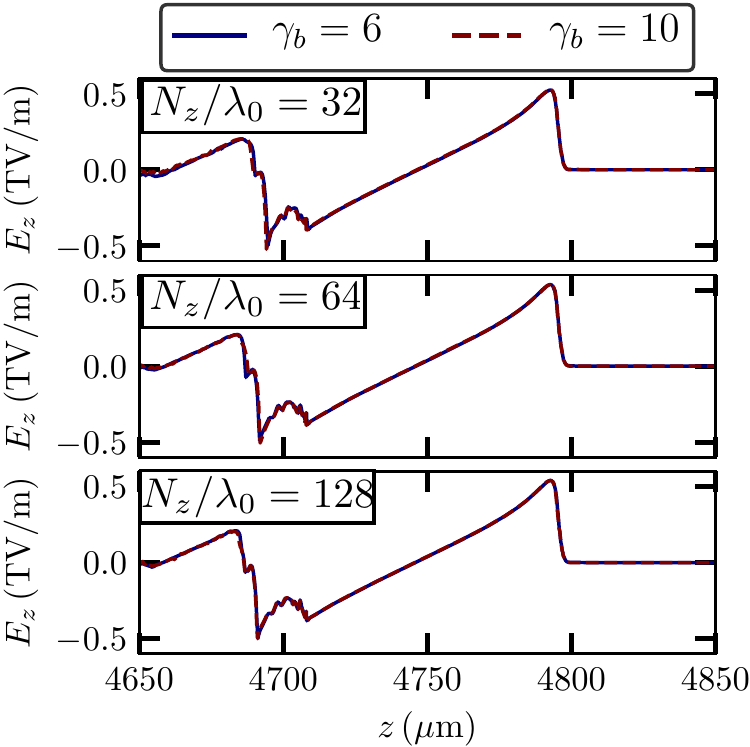}
\caption[A series of plots showing wakefield at $z=200\,\mathrm{\mu m}$ for the PSATD solver.]{A series of plots showing the wakefield when the laser reaches $z=4800\,\mathrm{\mu m}$. Each panel corresponds to a specific longitudinal resolution (given in the box on the upper left). Each panel shows the wakefield of 2-1/2D simulations using the CK solver carried out with $\gamma_b\in [6,10]$.}
\label{fig:field-layout-fdtd-1018}
\end{figure}

Simulations were conducted using the CK solver with the laser-plasma parameters given in Table~\ref{table:list-parameters}. A comparison of the simulated longitudinal accelerating wakefield on axis is shown in Fig.~\ref{fig:field-layout-fdtd-1018}, when the laser reaches $z=4800\,\mathrm{\mu m}$ using $\gamma_b\in [6,10]$ for $N_z/\lambda_0\in[32,64,128]$. We verify that, as expected, the plasma wavelength is elongated and the maximum amplitude is reduced, as compared with a higher plasma density case at $10^{19}\,\mathrm{cm^{-3}}$. The alteration of the accelerating wakefield at $\mathrm{4700\,\mu m}$ is a signature of beam loading effects. For each longitudinal resolution, a good agreement is obtained by comparing results from simulations at each $\gamma_b$.

\begin{figure}[htb]
\centering
\includegraphics{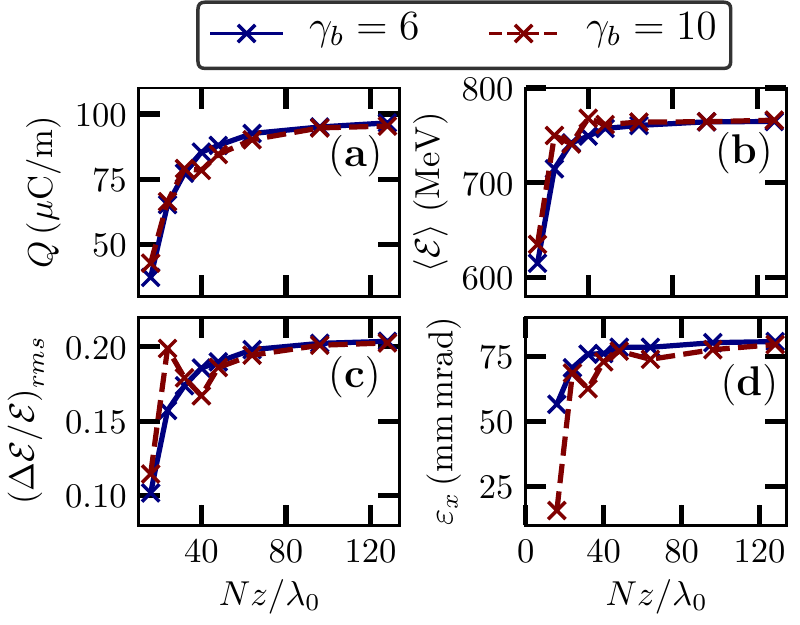}
\caption[Properties of injected and accelerated electron bunch for the PSATD solver using the Lorentz-boosted frame technique for the PSATD solver.]{Properties of injected and accelerated electron bunch evaluated at $z=4800\,\mu m$ with respect to the longitudinal resolution for $\gamma_b\in[2,3]$. Simulations were carried out using Warp in 2-1/2D using the PSATD solver with stencil order $32$. (a) Electron bunch charge $Q$, (b) average energy $\left<\mathcal{E}\right>$, (c) rms energy spread $(\Delta \mathcal{E}/\mathcal{E})_{rms}$, (d) transverse emittance with respect to longitudinal resolution $N_z/\lambda_0$.}
\label{fig:beam-prop-fdtd1018}
\end{figure}

Fig.~\ref{fig:beam-prop-fdtd1018} shows the beam properties with respect to the longitudinal resolution $N_z/\lambda_0$ at frame $z=4800\,\mathrm{\mu m}$. We have verified that other frames have shown the same tendency. All beam properties show a convergence as the longitudinal resolution increases. From $N_z/\lambda_0\geq64$, results from $\gamma_b=6$ and 10 converge to within $1\%$ of difference for the electron bunch charge, average energy, and energy spread. This result is only achieved at $N_z/\lambda_0>96$ for the transverse emittance.

\subsubsection{PSATD solver}
\label{sec:low-PSATD}

Simulations with the same laser-plasma parameters as in Section~\ref{sec:low-CK} were carried out using the PSATD solver. In this section, we report on the results of the convergence study of the longitudinal electric field and accelerated electron bunch properties with respect to the longitudinal resolution $N_z/\lambda_0$. 

\begin{figure}[htb]
\centering
\includegraphics{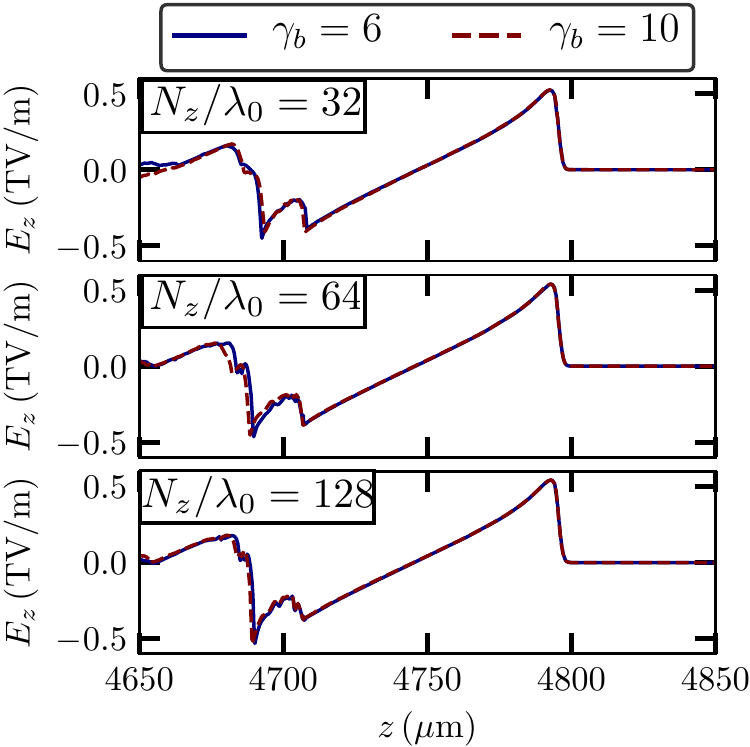}
\caption[A series of plots showing wakefield at $z=200\,\mathrm{\mu m}$ for the PSATD solver.]{Same as Fig.~\ref{fig:field-layout-fdtd-1018} but with the PSATD solver.}
\label{fig:spectral-layout-fdtd-1018}
\end{figure}

Fig.~\ref{fig:spectral-layout-fdtd-1018} shows the comparison of the longitudinal electric field from Lorentz-boosted frame simulations with $\gamma_b\in[6,10]$ taken when the laser reaches $z=4800\,\mathrm{\mu m}$. Each panel corresponds to a different longitudinal resolution, $N_z/\lambda_0\in[32,64,128]$. Results are very similar to the ones obtained with the CK solver, as shown in Fig.~\ref{fig:field-layout-fdtd-1018}. Some discrepancies can be observed at the back of the first period plasma wave with $N_z/\lambda_0=32,64$, but results at $N_z/\lambda_0=128$ show an excellent agreement.

\begin{figure}[htb]
\centering
\includegraphics{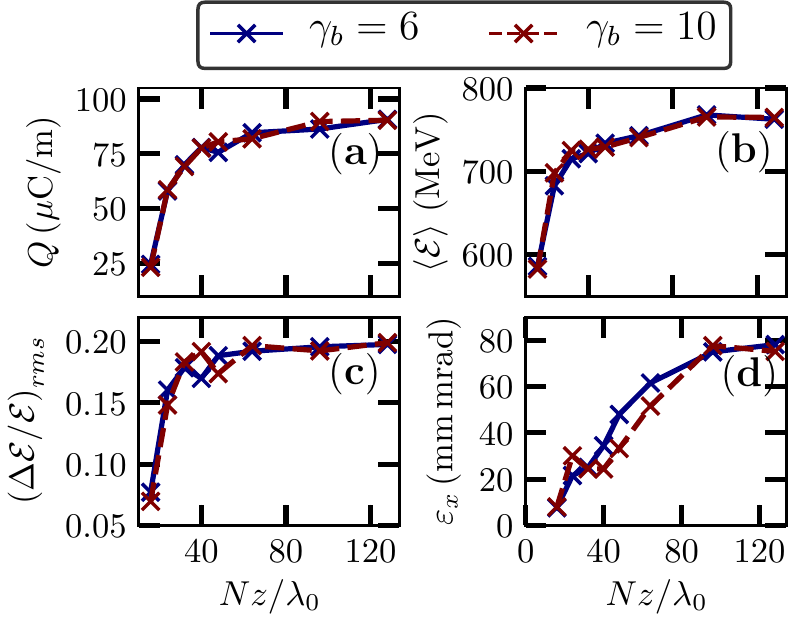}
\caption[Properties of injected and accelerated electron bunch for the PSATD solver using the Lorentz-boosted frame technique for the PSATD solver.]{Same as \ref{fig:beam-prop-fdtd1018} but with the PSATD solver.}
\label{fig:beam-prop-psatd1018}
\end{figure}

The convergence of the accelerated electron bunch properties with respect to $N_z/\lambda_0$ at the frame $z=4800\,\mathrm{\mu m}$ is reported in Fig.~\ref{fig:beam-prop-psatd1018}. The convergence is achieved for all beam properties at high longitudinal resolution, $N_z/\lambda_0\geq96$. At $N_z/\lambda_0=128$, the value of each beam quantity is within $5\%$ of the runs with the CK solver. 

\section{Runtime analysis}
\label{sec:runtime}
Comparisons of simulation runtimes give an insight on the speedup of the simulations performed using the Lorentz-boosted frame. All simulations were carried out on the Cray XC30 supercomputer Edison at the U.S. Department of Energy National Energy Research Supercomputer Center (NERSC) \cite{noauthor_national_2018}. 
In this analysis, the time for the diagnostics is subtracted from the total running time.

\begin{figure}[htb]
\centering
\includegraphics{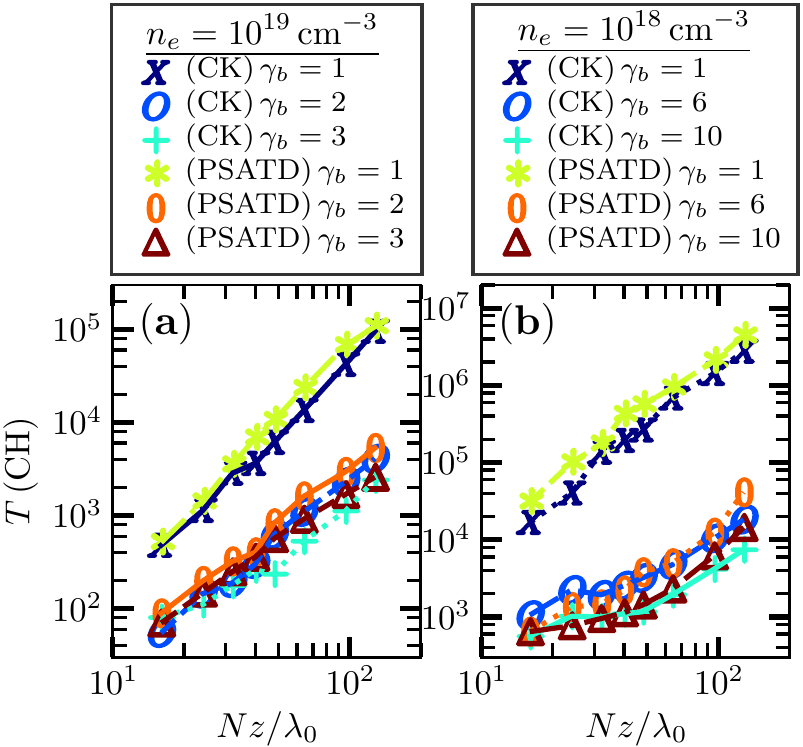}
\caption[Runtime of the simulations expressed in Core-Hours (CH) performed using Warp with respect to the longitudinal resolution.]{
Runtime of the simulations expressed in Core-Hours (CH) performed using Warp with respect to the longitudinal resolution, $N_z/\lambda_0$ for (a) $10^{19}\,\mathrm{cm^{-3}}$ and (b) $10^{18}\,\mathrm{cm^{-3}}$. Results from both the CK and the PSATD solvers are plotted. Note that $\gamma_b=1$ corresponds to the lab frame and that the plots are in log-scale.}
\label{fig:runtime}
\end{figure}

Fig.~\ref{fig:runtime} shows the runtime expressed in Core-Hours (CH) with respect to the longitudinal resolution $N_z/\lambda_0$. We observe that modeling LWFA in a $500\,\mathrm{\mu m}$ long plasma column with Warp using the CK solver in 2-1/2D in the laboratory frame at a longitudinal resolution ($N_z/\lambda_0=64$) requires $10^4\,$Core-Hours. To save computer time, complete simulations in the laboratory frame were only performed for the CK solver. In order to evaluate the runtime in the laboratory frame using the PSATD solver for each resolution, we ran the simulation up to 1000 steps and recorded its runtime. This obtained runtime was then used to extrapolate the runtime that would be required to model the full $500\,\mathrm{\mu m}$ plasma column, taking into account the observed nonlinear increase of the runtime (calibrated with the simulation time evolution given by the simulation using the CK solver, assuming similar nonlinear profile). 

When performing the same simulation in a boosted frame with $\gamma_b=3$, the computational cost is reduced by $\sim 20$ for the CK and the PSATD solvers, while retaining the difference within the percentage level as shown in Fig.~\ref{fig:beam-prop-fdtd-error}. As expected, simulations using the PSATD solver are more computationally expensive than the ones using the CK solver. Note that the PSATD solver that was available at the time of the study was a non-optimized prototype and the comparison of timing between the PSATD and CK runs given here are not meaningful. An optimized implementation of the PSATD solver is near completion and comparisons with optimized CK simulations will be reported elsewhere when available.


Runtime for the $1\,\mathrm{cm}-$LWFA modeling is given in Fig.~\ref{fig:runtime}(b). Complete simulations in the laboratory frame were not performed to completion in this case due to their high computational cost. The total runtime for those runs were extrapolated using the method described above in the laboratory frame using the PSATD solver in the $500\,\mathrm{\mu m}-$LWFA modeling. It is estimated that the runtime to model a $1\,\mathrm{cm}$ plasma is $\sim 1 \,\mathrm{ Million\,CH}$ in the laboratory frame at $N_z/\lambda_0=64$. Performing the same simulation using $\gamma=10$ reduces the runtime by $\sim 350$.

\begin{figure}[htb]
\centering
\includegraphics{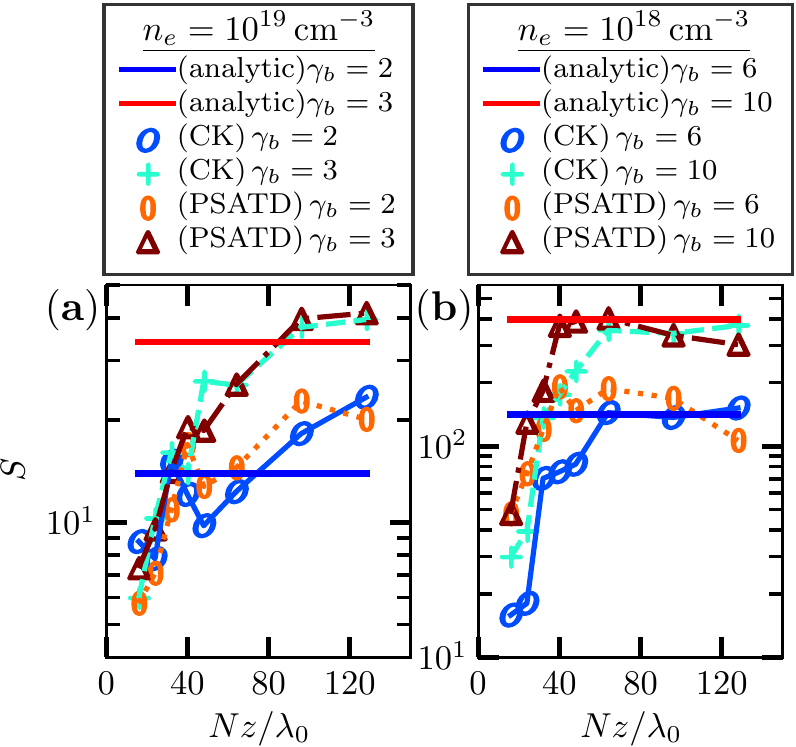}
\caption[Speedup with respect to the longitudinal resolution for the CK solver.]{Speedup with respect to the longitudinal resolution for (a) $10^{19}\,\mathrm{cm^{-3}}$ and (b) $10^{18}\,\mathrm{cm^{-3}}$. The analytical speedups are given by $S=\gamma_b^2\left(1+\beta_b\right)^2$ and the measured speedups are obtained from Warp simulations for both the CK and the PSATD solvers.
}
\label{fig:runtime-gain}
\end{figure}

The speedups from Warp simulations for the CK and the PSATD solvers at (a) $n_e = 10^{19}\,\mathrm{cm^{-3}}$ and (b) $n_e = 10^{18}\,\mathrm{cm^{-3}}$ are plotted In Fig.~\ref{fig:runtime-gain}, as well as the analytical speedup estimate $S=\gamma_b^2\left(1+\beta_b\right)^2$ \cite{vay_modeling_2011}. The speedup obtained with Warp simulations is of the same order of magnitude as the analytical estimate, and varies between 6 at lowest resolution with small $\gamma$ to 400 at high resolution for $\gamma=10$.  

\section{Conclusion}
We have performed convergence studies of LWFA stages at plasma densities of $10^{19}\,\mathrm{cm^{-3}}$, and $10^{18}\,\mathrm{cm^{-3}}$ in various Lorentz-boosted frames. The laser-plasma parameters were chosen such that the LWFA stages operate in the nonlinear regime with electron self-injection, triggered by a high $a_0$.

Simulations were performed using the finite-difference CK and the pseudo-spectral PSATD solvers. Results obtained demonstrated accurate modeling of the evolution of the plasma wakefield, electron bunch properties such as the charge, the average energy, the energy spread and the transverse dynamics with agreement at $99\,\%$ percentage level between simulations using various relativistic factors of the Lorentz-boosted frame, provided that the longitudinal resolution is sufficient, e.g. $N_z/\lambda_0=64$. The agreement between results using the CK and the PSATD solvers also agree at the $99\,\%$ percentage level at the highest longitudinal resolution $N_z/\lambda_0=128$ for both plasma densities. 


Speedups of one to two orders of magnitude were demonstrated on highly converged simulations with fluctuations on the various physical quantities of less than a percent. This validates that the Lorentz boosted frame method can be used to speedup significantly LWFA simulations in a highly non-linear regime with self-injection. Future work will explore higher energy stages at lower densities and extend to three-dimensions.

\section*{Acknowledgments}
We are thankful to Carlo Benedetti and Remi Lehe for insightful discussion. This work was supported by the Director, Office of Science, Office of High Energy Physics, of the U.S. Department of Energy under Contract No. DE-AC0205CH11231. This research used resources of the National Energy Research Scientific Computing Center, a DOE Office of Science User Facility supported by the Office of Science of the U.S. Department of Energy under Contract No. DE-AC02-05CH11231.

This document was prepared as an account of work sponsored in part by the United States Government. While this document is believed to contain correct information, neither the United States Government nor any agency thereof, nor The Regents of the University of California, nor any of their employees, nor the authors makes any warranty, express or implied, or assumes any legal responsibility for the accuracy, completeness, or usefulness of any in- formation, apparatus, product, or process disclosed, or represents that its use would not infringe privately owned rights. Reference herein to any specific commercial product, process, or service by its trade name, trademark, manufacturer, or otherwise, does not necessarily constitute or imply its endorsement, recommendation, or favoring by the United States Government or any agency thereof, or The Regents of the University of California. The views and opinions of authors expressed herein do not necessarily state or reflect those of the United States Government or any agency thereof or The Regents of the University of California.

\appendix*
\section{Electron bunch selection}
\label{app}
The electron bunch is selected following certain criteria:
\begin{itemize}
\item[$-$] electrons are situated in the first-period plasma wave, a region delimited by the zero-crossing of the wakefield,
\item [$-$] electrons are chosen above an arbitrary Lorentz factor, $\gamma_e$. 
\end{itemize}

\begin{figure}[htb]
\centering
\includegraphics{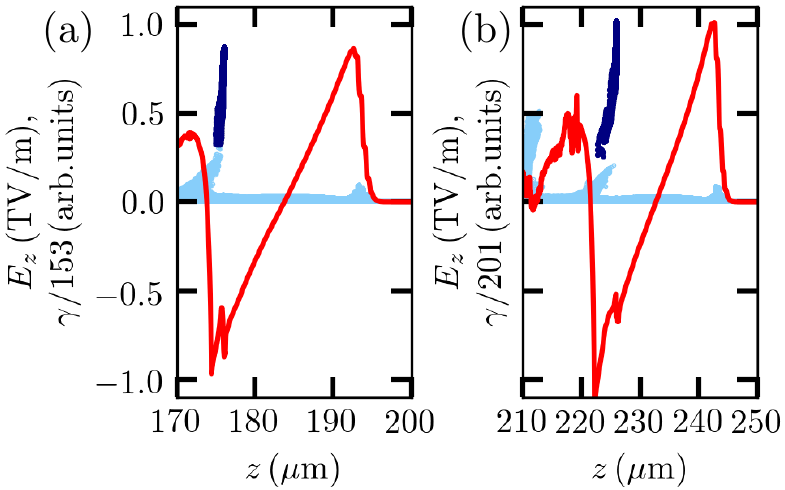}
\caption{Snapshots of longitudinal on-axis electric field $E_z$ (red) and electron distribution (light blue/light gray) represented in ($z,\gamma$) for the case at $n_e=10^{19}\,\mathrm{cm^{-3}}$, using the CK solver at $N_z/\lambda_0=64$. The electron bunch selected for analysis is in dark blue/dark gray. (a) Snapshot taken when the laser is approaching $200\,\mathrm{\mu m}$, and (b) approaching $250\,\mathrm{\mu m}$.}
\label{fig:selection}
\end{figure}

For illustration, we consider the case with plasma density $n_e=10^{19}\,\mathrm{cm^{-3}}$, using the CK solver at $N_z/\lambda_0=64$. $\gamma_e$ is fixed at $50$, which remained unchanged while evaluating the evolution of beam properties. In Fig.~\ref{fig:selection} are shown two snapshots of the longitudinal electric field on axis, $E_z$, and the electron distribution represented in ($z,\gamma$) in light blue/light gray. The selected electrons are in the blue/gray region, which is within the first-period plasma wave, and above $\gamma_e=50$. This threshold value allows a clear separation of the accelerated electron bunch from the blob for these two snapshots, however it is not guaranteed beyond these snapshots, nevertheless we still achieve a convergence as shown in Fig.~\ref{fig:beam-prop-evol-fdtd} and Fig.~\ref{fig:beam-spectral-evol}. For the study at $n_e=10^{18}\,\mathrm{cm^{-3}}$, we also observe a separation of an electron bunch from the blob by fixing $\gamma_e=1000$.

\bibliographystyle{apsrev4-1}
\bibliography{biblio}
\end{document}